\documentclass[12pt]{article}
\usepackage{amsmath,latexsym}
\usepackage{graphicx}
\begin{document}

 \title{\Huge Thin shell wormholes in higher dimensiaonal Einstein-Maxwell theory}
 \author{F.Rahaman$^*$, M.Kalam$^{\ddag}$ and  S.Chakraborty$^\dag$  }
\date{}
 \maketitle

 \begin{abstract}
We construct thin shell Lorentzian wormholes in higher
dimensional Einstein-Maxwell theory applying the ' Cut and Paste
' technique proposed by Visser. The linearized stability is
analyzed under radial perturbations around some assumed higher
dimensional spherically symmetric static solution of the Einstein
field equations in presence of Electromagnetic field. We
determine the total amount of exotic matter, which is concentrated
at the wormhole throat.
\end{abstract}

 %\bigskip
 %\medskip
  \footnotetext{ Pacs Nos :  04.20 Gz,04.50 + h, 04.20 Jb   \\
 Key words:  Thin shell wormholes , Electromagnetic field,  Higher Dimension, Stability
\\
 $*$Dept.of Mathematics, Jadavpur University, Kolkata-700 032,
 India:
                                  E-Mail:farook\_rahaman@yahoo.com\\
$\ddag$Dept. of Phys. , Netaji Nagar College for Women, Regent Estate, Kolkata-700092, India.\\
  $\dag$Dept. of Maths., Meghnad Saha Institute of Technology,
                                           Kolkata-700150, India
}
    \mbox{} \hspace{.2in}

\title{\Huge Introduction: }

In a pioneer work, Morris and Thorne [1] have  found traversable
Lorentzian wormholes as the solutions of Einstein's field
equations. These are hypothetical shortcuts between two regions (
of the same Universe or may be of two separate Universes )
connected by a throat. The throat of the wormholes is defined as
a two dimensional hypersurface of minimal area and to hold such a
wormhole open, violations of certain energy conditions are
unavoidable i.e. the energy momentum tensor of the matter source
of gravity violates the local and averaged null energy condition
$ T_{\mu\nu} k^\mu k^\nu \geq 0 $, $ k_\mu k^\nu = 0 $ . Thus all
traversable wormholes require exotic matter that violates the
null energy condition. Recently, it has been shown that the
requirements of exotic matter for the existence of a wormhole can
be made infinitesimally small by  a suitable choice of the
geometry [2-3].

\pagebreak

 In recent past, Visser [4] has proposed
another way, which is known as 'Cut and paste' technique,  of
minimizing the usage of exotic matter to construct a wormhole in
which the exotic matter is concentrated at the wormhole throat.
In 'Cut and paste' technique, the wormholes are theoretically
constructed by cutting and pasting two manifolds to obtain
geodesically complete new manifold with a throat placed in the
joining shell [5]. Using Darmois-Israel [6] formalism, one can
determine the surface stresses of the exotic matter ( located in
thin shell placed at the joining surface ). Though we do not know
about the equation of state of exotic matter, yet it is possible
to investigate the stability of these thin wormholes. Following
references [5-6], one can analyze  the stability of these thin
wormholes through linearized perturbations around static
solutions of the Einstein field equations. Several authors have
used surgical technique (Cut and Paste) to construct thin
wormholes. Poisson and Visser have analyzed the stability of a
thin wormhole constructed by joining two Schwarzschild spacetimes
[5].  Eiroa and Romero  [7] have extended the linearized
stability analysis to Reissner-Nordstr\"{o}m thin spacetimes.
Eiroa and Simeone have constructed the wormholes by cutting and
pasting two metrics corresponding to a charged black hole which
is a solution of low energy bosonic string theory, with vanishing
antisymmetric field but including a Maxwell field[8]. Also the
same authors have analyzed cylindrically symmetric thin wormhole
geometry associated to gauge cosmic strings [9]. Recently,
Thibeault et al [10] have studied the stability and energy
conditions of five dimensional spherically symmetric thin shell
wormholes in Einstein-Maxwell theory with addition of a Gauss
Bonnet term. In this article, we study thin shell wormholes in
higher dimensional Einstein-Maxwell theory i.e. wormholes
constructed by cutting and pasting two metrics corresponding a
higher dimensional Reissner-Nordstr\"{o}m black hole. We are
interested only to study the geometry of these objects. We do not
explain about the mechanism that provide the exotic matter to
them, but rather we focus on the total amount of exotic matter.

\title{\Huge2. Reissner-Nordstr\"{o}m Black holes in higher dimension: }

The  Reissner-Nordstr\"{o}m Black hole is a solution of the
Einstein equation coupled to the Maxwell field. From the
Einstein-Maxwell action in ( D+2) dimension [11]

\begin{equation}
              S
             =\int d^{D+2} \sqrt{-g} [ R - \frac{k}{8\pi} F_{ab}
             F^{ab}]
           \end{equation}

where
  \begin{equation}k = 8\pi G    \end{equation}
\begin{equation}F_{ab} = A_{a;b} -  A_{b;a}   \end{equation}

\pagebreak

One can obtain the following Einstein-Maxwell equations

 \begin{equation}
              R_{ab} - \frac{1}{2}g_{ab}R
             = \frac{k}{4\pi} [F_{a}^{c} F_{bc} - \frac{1}{4}g_{ab}F_{cd}
             F^{cd}]
           \end{equation}

\begin{equation}F_{a;c}^{c} = 0    \end{equation}

\begin{equation}F_{ab;c} + F_{bc;a} + F_{ca;b} = 0    \end{equation}

The only non trivial components of $F_{ab}$ are

\begin{equation} F_{tr} = - F_{rt} = \frac{Q}{r^D} \end{equation}

where Q represents an isolated point charge.

These equations admit a spherically symmetric static solution
given by [11]

\begin{equation}
               ds^2=  - f(r) dt^2+ \frac{dr^2}{ f(r)} + r^2 d\Omega_D^2
               \end{equation}

where $ d\Omega_D^2 $ is the line element on the D unit sphere
i.e.

\begin{equation}
                d\Omega_D^2 =  d{\theta}_1^2 + \sin^2{\theta}_1 d
                {\theta}_2^2+ ...... + \prod_{n=1}^{D-1}\sin^2
                {\theta}_n d{\theta}_D^2
               \end{equation}
The volume of the D unit sphere is given by

\begin{equation}
                \Omega_D =  2\frac{ \pi
                ^\frac{D+1}{2}}{\Gamma(\frac{D+1}{2})}
               \end{equation}
The expression of f(r) is  [11]

\begin{equation}
           f(r) = 1 -  \frac{ \mu}{r^{D-1}} + \frac{ q^2}{r^{2(D-1)}}
               \end{equation}

\pagebreak

 Here, $\mu$ is related to the mass M of the black hole
as

\begin{equation}
           \mu =  \frac{16 \pi GM }{D\Omega_D}
               \end{equation}
\begin{equation}
          q^2 =  \frac{8 \pi GQ }{D(D-1)}
               \end{equation}

There are two roots of equation $ f(r) = 0 $ as

\begin{equation}
           r_{\pm} = [ \frac{\mu}{2} \pm \frac{\mu}{2}( 1 -
           \frac{4q^2}{\mu^2}) ]^{ \frac{1}{D-1}}
               \end{equation}

When $ \mu^2 < 4q^2 $, we have two positive roots, one of which
is an outer horizon $r_+$ while the other is inner horizon.

If $ \mu^2 = 4q^2 $, both horizons coincide at
\begin{equation}
           r_+ = r _- = [ \frac{8 \pi GM }{D\Omega_D}]^{ \frac{1}{D-1}}
               \end{equation}

\title{\Huge3. The Darmois-Israel formalism
                and Cut and Paste construction: }

From the higher dimensional Reissner-Nordstr\"{o}m geometry, we
can take two copies of the region with $ r\geq a$ :

$ M^\pm = ( x \mid r \geq a )  $

and paste them at the hypersurface

$ \Sigma = \Sigma^\pm = ( x \mid r = a )  $

We take $ a > r_+ $ to avoid horizon and this new construction
produces a geodesically complete manifold $ M = M^+ \bigcup M^- $
with a matter shell at the surface $ r = a $ , where the throat
of the wormhole is located. Thus M is a manifold with two
asymptotically flat regions connected by the throat. We shall use
the Darmois-Israel formalism to determine the surface stress at
the junction boundary. Now we choose the coordinates $ \xi^i (
\tau, {\theta}_1, {\theta}_2, ........, {\theta}_D)$ in $\Sigma$
where the throat is located with $\tau$ is the proper time on the
shell.

\pagebreak

 To analyze the dynamics of the wormhole, we let the
radius of the throat be a function of the proper time $ a =
a(\tau)$.

The parametric equation for $\Sigma$ is given  by

\begin{equation}\Sigma : F(r,\tau ) = r - a(\tau)\end{equation}

The intrinsic surface stress energy tensor, $ S_{ij} $ is given
by the Lanczos equation in the following form

\begin{equation}S_j^i  =  - \frac{1}{8\pi} ( \kappa_j^i  - \delta_j^i \kappa_k^k )\end{equation}

where $ \kappa_{ij} = K_{ij}^+ - K_{ij}^- $ i.e. the discontinuity
in the second fundamental forms or extrinsic curvatures.

The extrinsic curvature associated with the two sides of the
shell are

\begin{equation}K_{ij}^\pm =  - n_\nu^\pm\ [ \frac{\partial^2X_\nu}
{\partial \xi^i\partial \xi^j } +
 \Gamma_{\alpha\beta}^\nu \frac{\partial X^\alpha}{\partial \xi^i}
 \frac{\partial X^\beta}{\partial \xi^j }] |_\Sigma \end{equation}

where $ n_\nu^\pm\ $ are the unit normals to $\Sigma$,

\begin{equation} n_\nu^\pm =  \pm   | g^{\alpha\beta}\frac{\partial F}{\partial X^\alpha}
 \frac{\partial F}{\partial X^\beta} |^{-\frac{1}{2}} \frac{\partial F}{\partial X^\nu} \end{equation}

with $ n^\mu n_\mu = 1 $.

The intrinsic metric on $\Sigma$ is given by

\begin{equation}
               ds^2 =  - d\tau^2 + a(\tau)^2 d\Omega_D^2
               \end{equation}

\pagebreak

 From Lanczos equation, one obtain the surface stress
energy tensor $ S_j^i = diag ( - \sigma , p_{\theta_1},
p_{\theta_2}, ...., p_{\theta_D}) $

where $ \sigma$ is the  surface  energy density and p is the
surface pressure as

\begin{equation}
               \sigma =  - \frac{D}{4\pi a}\sqrt{f + \dot{a}^2}
               \end{equation}

\begin{equation}
              p_{\theta_1} = p_{\theta_2} =
.... =  p_{\theta_D} = p =  - \frac{D-1}{D}\sigma + \frac{1}{8\pi
}\frac{2\ddot{a} + f^\prime }{\sqrt{f + \dot{a}^2}}
               \end{equation}

where over dot and prime mean, respectively, the derivatives with
respect to $\tau$ and r.

From equations (21) and (22), one can verify the energy
conservation equation:

\begin{equation}
               \frac {d}{d \tau} (\sigma a^D) + p \frac{d}{d \tau}(a^D)= 0
               \end{equation}
or
\begin{equation}
               \dot{\sigma} + D \frac{\dot{a}}{a}( p + \sigma ) = 0
               \end{equation}

The first term represents the variation of the internal energy of
the throat and the second term is the work done by the throat's
 internal forces. Negative energy density in (21) implies  the
existence of exotic matter at the shell.

\title{\Huge4. Linearized Stability Analysis: }

Rearranging equation (21), we obtain the thin shell's  equation of
motion

            \begin{equation}  \dot{a}^2 + V(a)= 0  \end{equation}

                Here  the potential is defined  as
\begin{equation}
              V(a) =  f(a) - \frac{16\pi^2 a^2\sigma^2(a)}{D^2}
                 \end{equation}

\pagebreak

 Linearizing around a static solution situated at $a_0$,
one can expand V(a) around $a_0$ to yield

\begin{equation}
              V =  V(a_0) + V^\prime(a_0) ( a - a_0) + \frac{1}{2} V^{\prime\prime}(a_0)
              ( a - a_0)^2 + 0[( a - a_0)^3]
                 \end{equation}

where prime denotes derivative with respect to $a$.

Since we are linearizing around a static solution at $ a = a_0 $,
we have $ V(a_0) = 0 $ and $ V^\prime(a_0)= 0 $. The stable
equilibrium configurations correspond to the condition $
V^{\prime\prime}(a_0)> 0 $. Now we define a parameter $\beta$,
which is interpreted as the speed of sound, by the relation
\begin{equation}
              \beta^2(\sigma) = \frac{ \partial p}{\partial
              \sigma}|_\sigma
                 \end{equation}
Using conservation equation (24), we have
\begin{equation} V^{\prime\prime}(a) = f^{\prime\prime} - \frac{32\pi^2 a^2\sigma^2}{D^2}
  - \frac{128\pi^2 a \sigma \sigma^\prime}{D^2}   - \frac{32\pi^2 a^2 (\sigma^{\prime})^2}{D^2}
 - \frac{32\pi^2 a^2\sigma}{D^2} [ \frac{D}{a^2} ( p + \sigma ) -
 \frac{D}{a} \sigma^{\prime} ( 1 + \beta^2) ] \end{equation}
The wormhole solution is stable if $ V^{\prime\prime}(a_0)> 0 $
i.e. if
\begin{equation}
              \beta_0^2 < \frac{ 1}{D(a_0f_0^\prime - 2 f_0 )}[
              a_0f_0^\prime- 2f_0 - a_0^2 f_0^{\prime\prime} +
              \frac{a_0^2(f_0^\prime)^2}{2f_0}] - 1
                 \end{equation}
or
\begin{equation}
              \beta_0^2 < \frac{ 1}{D} - 1 + \frac {[\frac{2\mu(D-1)(D-2)}{a_0^{D-1}}+
              \frac{\mu^2(D-1)(9D-15)}{a_0^{2D-2}}-\frac{2\mu q^2(D-1)(4D-7)}{a_0^{3D-3}}+
              \frac{4q^4(D-1)(D-2)}{a_0^{4D-4}}]}{2(1 -  \frac{\mu}{a_0^{D-1}} +
              \frac{q^2}{a_0^{2(D-1)}})(2 -  \frac{\mu (D+1)}{a_0^{D-1}} +
              \frac{2q^2D}{a_0^{2(D-1)}})}
                 \end{equation}
Thus if one treats $a_0$, D and the parameters related to the
Reissner-Nordstr\"{o}m black hole are specified quantities, then
the stability of the configuration requires the above restriction
on the parameter $\beta_0$.

\title{\Huge5. Energy condition and exotic matter: }

Weak energy condition( WEC ) implies that for all time like
vectors $x^\mu$, $ T_{\mu\nu}x^\mu x^\nu \geq 0 $. In an
orthonormal basis WEC reads $ \rho \geq 0, \rho + p_i \geq 0
\forall i$, where $\rho$ is the energy density and $p_i$ , the
principal pressures. Null energy condition ( NEC ) states that $
T_{\mu\nu}k^\mu k^\nu \geq 0 $ for all null vectors $k^\mu$. In
an orthonormal frame $ T_{\mu\nu}k^\mu k^\nu \geq 0 $ takes the
form $ \rho + p_i \geq 0 \forall i$. [ The WEC implies by
continuity the NEC ]. In the case of thin wormhole constructed
above, we have ( from equations (21) and (22) ) $ \sigma < 0 $
and $ \sigma + p  < 0 $ i.e. matter occupies in the shell
violates WEC and NEC, in other words, shell contains exotic
matter. The only contributor in the stress tensor out side the
shell is electromagnetic field. Now from the field equations $
R_{ab} - \frac{1}{2}g_{ab}R = 8\pi G T_{ab} $, we can write $
T_{ab} = T_{ab}^{EM}
             = \frac{1}{4\pi} [F_{a}^{c} F_{bc} - \frac{1}{4}g_{ab}F_{cd}
             F^{cd}]$.

Now, the energy density $ \rho = T_{tt}$ , the radial pressure $
p_R = T_{rr}$ and the tangential pressure $ p_t = p _\theta = p
_\phi $ are given by $ \rho^{EM} = p_t^{EM} = - p_r^{EM} =
\frac{q^2}{r^{2D}} $. Thus $ \rho^{EM} > 0 $ , $ \rho^{EM} +
p_t^{EM}  > 0 $  and $ \rho^{EM} + p_r^{EM} = 0 $ i.e. the NEC
and WEC are satisfied out side the shell. Hence the exotic matter
is confined within the shell. The total amount of exotic matter
can  be quantified by the integrals [3]

$\int \rho\sqrt{-g}d^{D+1}x $ , $\int [\rho + p_i]
\sqrt{-g}d^{D+1}x $ , where g is the determinant of the metric
tensor. To quantify the amount of exotic matter, we use the
following integral ( NEC violating matter is related only on
$p_r$ and not the transverse components )
\begin{equation}
             \Omega =  \int [\rho + p_r]
\sqrt{-g}d^{D+1}x
                 \end{equation}

Following Eiroa and Simone [8] , we introduce new radial
coordinate $ R  =  \pm ( r -a ) $ in M ( $\pm $ for $M^{\pm}$
respectively ) as

\begin{equation}
            \Omega =  \int_0^{2\pi} \int_0^\pi\ ....\int_0^\pi \int_{-\infty}^\infty [\rho + p_r]
\sqrt{-g}dRd{\theta_1}d{\theta_2}.....d{\theta_D}
                 \end{equation}

Since the shell does not exert radial pressure and the energy
density is located on a thin shell surface, so that $ \rho =
\delta(R)\sigma_0$, then we have

$
             \Omega = \int_0^{2\pi} \int_0^\pi ....\int_0^\pi [\rho
\sqrt{-g} ]|_{r=a_0} d{\theta_1}d{\theta_2}.....d{\theta_D}$

= $a_0^D\sigma_0\times $ area of the unit D-sphere $ $

 = $ 2 a_0^D\sigma_0
\frac{ \pi^\frac{D+1}{2}}{\Gamma(\frac{D+1}{2})} $

Thus one gets,
\begin{equation}
            \Omega =  - D a_0^{D-1}\sqrt{f_0}
\frac{ \pi^\frac{D+1}{2}}{2\Gamma(\frac{D+1}{2})}
                 \end{equation}

                 Using eq.(10), we have
\begin{equation}
            \Omega =  - D a_0^{D-1}\sqrt{1 -  \frac{ \mu}{a_0^{D-1}} + \frac{ q^2}{a_0^{2(D-1)}}}
\frac{ \pi^\frac{D+1}{2}}{2\Gamma(\frac{D+1}{2})}
                 \end{equation}

Since the total amount of exotic matter $ \Omega$ is proportional
to $\sqrt{f_0}$, then $ \Omega$ approaches to zero when wormhole
radius tends to the event horizon ( i.e. when $ a_0 \rightarrow
r_+  $ ). So one can get vanishing amount of exotic matter by
taking $a_0 $ near $r_+ $.

\pagebreak

\title{\Huge5. Concluding remarks: }

Recently,  several theoretical physicists are interested to
obtain wormholes by surgically grafting two identical copies of
various well known spacetimes. In this report, we have
constructed thin wormhole in higher dimensional Einstein-Maxwell
theory. We analyze the dynamical stability of the thin shell,
considering linearized radial perturbations around stable
solutions. To analyze this, we define a parameter $ \beta^2 =
\frac{p^\prime }{\sigma^\prime} $ as a parametrization of the
stability of equilibrium. We have obtained a restriction on
$\beta^2$ to get  stable equilibrium of the thin wormhole( see
eq.(31)).  We have shown that matter within the shell violates
the WEC and NEC but matter out side the shell obeys the NEC and
WEC. Thus the exotic matter is confined only within the shell.
Since the viability of traversable wormholes are linked to the
total amount of exotic matter for their construction, we have
calculated an integral measuring of the total amount of exotic
matter. Finally,  we have shown that total amount of exotic
matter needed to support traversable wormhole can be made
infinitesimal small by taking wormhole radius near the throat.

\pagebreak

 {  \Huge Acknowledgments }

          F.R. is thankful to Jadavpur University and DST , Government of India for providing
          financial support under Potential Excellence and Young
          Scientist scheme . MK has been partially supported by
          UGC,
          Government of India under Minor Research Project scheme. \\
We are thankful to the anonymous referees for their
 constructive suggestions.

%\begin{figure}[p]
%\includegraphics*[450,350]{fig1.bmp}
%\caption{Variation of deflection of the circular plate}
%\end{figure}


\begin{thebibliography}{99}
\bibitem{kg6}  M. Morris and K. Throne , American  J. Phys. 56, 39 (1988 )
\bibitem{kg10} M Visser, S Kar and N Dadhich
 Phys. Rev.Lett. 90, 201102(2003)[arXiv: gr-qc /
    0301003]
        \bibitem{kg10} P Kuhfittig Phys.Rev.D68:067502,2003
[ arXiv: gr-qc / 0401048 ]

\bibitem{kg10} M Visser Nucl.Phys.B 328, 203 ( 1989)
\bibitem{kg10} E Poisson and M Visser
Phys.Rev.D 52, 7318 (1995) [arXiv: gr-qc / 9506083]
\bibitem{kg10} W Israel Nuovo Cimento 44B , 1 (1966) ; erratum -
ibid. 48B, 463 (1967); F Lobo and P Crawford Class.Quan.Grav. 21,
391 (2004) ; M Ishak and K Lake Phys.Rev.D 65, 044011 (2002)
\bibitem{kg10}  E Eiroa and G Romero
Gen.Rel.Grav. 36, 651 (2004)[arXiv: gr-qc / 0303093]
\bibitem{kg10}  E Eiroa and C Simeone
Phys.Rev.D 71, 127501 (2005) [arXiv: gr-qc / 0502073]
\bibitem{kg10}  E Eiroa and C Simeone  Phys.Rev.D 70, 044008
(2004)
\bibitem{kg10}  M Thibeault , C Simeone and E Eiroa  arXiv: gr-qc /
0512029
\bibitem{kg10}  R Myers and M Perry Ann. Phys. ( N.Y.) 172, 304
(1986)


\end{thebibliography}
\end{document}